\documentclass[11pt]{article}

\usepackage{latexsym,amsmath,amssymb,amscd,eucal}
\usepackage{xypic}
\usepackage{enumerate}

\def\harr#1#2{\smash{\mathop{\hbox to .3in{\rightarrowfill}}
 \limits^{\scriptstyle#1}_{\scriptstyle#2}}}

\def\appendix#1{\addtocounter{section}{1}\setcounter{equation}{0}
\renewcommand{\thesection}{\Alph{section}}
\section*{Appendix \thesection\protect\indent \parbox[t]{11.715cm} {#1}}
\addcontentsline{toc}{section}{Appendix \thesection\ \ \ #1} }
\newcommand{\eq}{\begin{equation}}
\newcommand{\eqend}{\end{equation}}

\newbox\ncintdbox \newbox\ncinttbox
\setbox0=\hbox{$-$} \setbox2=\hbox{$\displaystyle\int$}
\setbox\ncintdbox=\hbox{\rlap{\hbox to
\wd2{\hskip-.125em\box2\relax \hfil}}\box0\kern.1em}
\setbox0=\hbox{$\vcenter{\hrule width 4pt}$}
\setbox2=\hbox{$\textstyle\int$}
\setbox\ncinttbox=\hbox{\rlap{\hbox to
\wd2{\hskip-.175em\box2\relax \hfil}}\box0\kern.1em}

\def\be{\begin{equation}}
\def\ee{\end{equation}}
\def\bea{\begin{eqnarray}}
\def\eea{\end{eqnarray}}
\def\bd{\begin{displaymath}}
\def\ed{\end{displaymath}}

\DeclareFontFamily{U}{rsf}{}
\DeclareFontShape{U}{rsf}{m}{n}{
  <5> <6> rsfs5 <7> <8> <9> rsfs7 <10-> rsfs10}{}
\DeclareMathAlphabet\Scr{U}{rsf}{m}{n}

\def\cR{{\Scr R}}

\def\cB{{\Scr B}}

\def\cH{{\Scr H}}

\def\cG{{\Scr G}}

\makeatletter
\newdimen\normalarrayskip              
\newdimen\minarrayskip                 
\normalarrayskip\baselineskip
\minarrayskip\jot
\newif\ifold             \oldtrue            
\def\arraymode{\ifold\relax\else\displaystyle\fi} 
\def\@arrayskip{\ifold\baselineskip\z@\lineskip\z@
     \else
     \baselineskip\minarrayskip\lineskip2\minarrayskip\fi}
\def\@arrayclassz{\ifcase \@lastchclass \@acolampacol \or
\@ampacol \or \or \or \@addamp \or
   \@acolampacol \or \@firstampfalse \@acol \fi
\edef\@preamble{\@preamble
  \ifcase \@chnum
     \hfil$\relax\arraymode\@sharp$\hfil
     \or $\relax\arraymode\@sharp$\hfil
     \or \hfil$\relax\arraymode\@sharp$\fi}}
\def\@array[#1]#2{\setbox\@arstrutbox=\hbox{\vrule
     height\arraystretch \ht\strutbox
     depth\arraystretch \dp\strutbox
     width\z@}\@mkpream{#2}\edef\@preamble{\halign \noexpand\@halignto
\bgroup \tabskip\z@ \@arstrut \@preamble \tabskip\z@ \cr}%
\let\@startpbox\@@startpbox \let\@endpbox\@@endpbox
  \if #1t\vtop \else \if#1b\vbox \else \vcenter \fi\fi
  \bgroup \let\par\relax
  \let\@sharp##\let\protect\relax
  \@arrayskip\@preamble}
\makeatother

\newcommand{\beq}{\begin{eqnarray}}
\newcommand{\eeq}{\end{eqnarray}}

\def\appendix#1{\addtocounter{section}{1}\setcounter{equation}{0}
\renewcommand{\thesection}{\Alph{section}}
\section*{Appendix \thesection. #1}
\protect\indent \parbox[t]{11.715cm}
\addcontentsline{toc}{section}{Appendix \thesection\ \ \ #1} }

\newtheorem{theorem}{Theorem}[section]

\numberwithin{equation}{section}

\begin{document}


\vspace{.1in}

\begin{center}
{\Large\bf AdS Solutions in Gauge Supergravities and the  \\ Global Anomaly for the Product of Complex Two-Cycles}

\end{center}
\vspace{0.1in}

\begin{center}
{\large
A. A. Bytsenko }$^{(a),}$\footnote{abyts@uel.br}
{\large and E. Elizalde }$^{(b),}$\footnote{elizalde@ieec.uab.es, elizalde@math.mit.edu}
\vspace{5mm}
\\
$^{(a)}$
{\it Departamento de F\'{\i}sica, Universidade Estadual de
Londrina,\\
Caixa Postal 6001, Londrina-Paran\'a, Brazil}
\vspace{3mm}\\
$^{(b)}$ {\it Consejo Superior de Investigaciones Cient\'{\i}ficas (ICE/CSIC)\\
and Institut d'Estudis Espacials de Catalunya (IEEC)\\
Campus UAB, Facultat de Ci\`{e}ncies, Torre C5-Par-2a pl \\
E-08193
Bellaterra (Barcelona) Spain}\\
\end{center}
\vspace{0.1in}

\begin{center}
{\bf Abstract}
\end{center}
{\small Cohomological methods are applied for the special set of solutions corresponding to
rotating branes in arbitrary dimensions, AdS black holes (which can be embedded in ten or eleven dimensions), and gauge supergravities. A new class of solutions is proposed, the Hilbert modular varieties, which consist of the $2n$-fold product of the
two-spaces ${\bf H}^n/\Gamma$ (where ${\bf H}^n$ denotes the product of $n$ upper half-planes, $H^2$, equipped with the co-compact action of $\Gamma \subset SL(2, {\mathbb R})^n$) and $({\bf H}^n)^*/\Gamma$ (where $(H^2)^* = H^2\cup \{{\rm cusp\,\, of}\,\,\Gamma\}$ and $\Gamma$ is a congruence subgroup of $SL(2, {\mathbb R})^n$).
The cohomology groups of the Hilbert variety, which inherit a Hodge structure (in the sense of Deligne), are analyzed, as well as bifiltered sequences, weight and Hodge filtrations, and it is argued that the torsion part of the cuspidal cohomology is involved in the global anomaly condition. Indeed, in presence of the cuspidal part, all cohomology classes can be mapped to the boundary of the space and the cuspidal contribution can be involved in the global anomaly condition.}

\vfill



\newpage


\section{Introduction}

Owing to the important and very useful correspondence that exists between the anti-de Sitter (AdS) space and conformal field theories on its boundary, AdS black hole solutions of gauged extended supergravities are currently attracting a good deal of attention. In particular, one should mention the celebrated Cardy-Verlinde formula relating the entropy of a conformal field in arbitrary dimensions to its total energy and Casimir energy \cite{cv1} (a general review of braneworld AdS theory in relation with the Cardy-Verlinde
formula is done in the third reference of \cite{cv1}). Also, the AdS/CFT Casimir energy for black holes is being specifically considered \cite{gpp1}.

Interesting in this context is to note that gauged extended supergravities can arise as the massless modes of various Kaluza-Klein compactifications of $D=11$ and $D=10$ supergravities. Gauged supergravity theories may be obtained by a consistent truncation of the massive modes of full Kaluza-Klein theories. It thus follows that all solutions of the lower-dimensional theories will also be solutions of the higher-dimensional ones. Specifically, the general rotating brane solutions in arbitrary dimensions, supported by a single charge, and their sphere reductions, have been discussed in \cite{Cvetic99}.

Rotating branes can be constructed by performing standard diagonal dimensional oxidations of general rotating black holes. In particular, one can consider the sphere reduction of generic rotating branes and associated domain wall black holes
(see, for details, \cite{Cvetic99}). Note that, while for asymptotically Minkowskian black holes the horizons are spherical, AdS black holes can admit horizons with some more general topology.

In this paper we propose a new class of solutions, the Hilbert modular varieties, which consists of the $2n$-fold product of the two-spaces ${\bf H}^n/\Gamma$
(where ${\bf H}^n$ denotes the product of $n$ upper half-planes $H^2$ equipped with the co-compact action of $\Gamma \subset SL(2, {\mathbb R})^n$)
and $({\bf H}^n)^*/\Gamma$ (where $(H^2)^* = H^2\cup \{{\rm cusp\,\, of}\,\,\Gamma\}$ and $\Gamma$ is a congruence subgroup of $SL(2, {\mathbb R})^n$).

This class of solutions can also come from type IIB and eleven-dimensional supergravity solutions,\footnote{Such class may appear as K\"ahler-Einstein space factors in supergravity/superstring vacua \cite{Bytsenko05,Bytsenko09}.}
and also in the eleven-dimensional supergravities warped product of AdS$_3$ with an eight-dimensional polynomial solution, which are dual to conformal field theories with $N=(0,2)$ supersymmetry (owing to the AdS-CFT correspondence).
For $n=1$, the embedding procedure considered in \cite{Cvetic99} yields that AdS compactifications are singled out, as arising from the near-horizon hyperbolic geometry of a two-brane rotating in the extra dimensions. In addition, the four charges corresponding to the $U(1)^{4}$ Cartan subgroup are just the four angular momenta.
The class of our K\"{a}hler-Einstein factorized solutions (and their anz\"{a}tze) is a new type of the IIB and eleven-dimensional bubble solutions \cite{Kim1,Kim2}. This type of $2n$-fold product also can be derived from the general three-charged and four-charged superstar solutions of the type IIB and eleven-dimensional supergravity, respectively. (Note that the three- and four-charge superstar geometry has been described in \cite{Cvetic1}.)

We will consider in detail, in Sect.~2 the Hilbert modular variety, its cohomology groups, which inherit a Hodge structure (in the sense of Deligne), bifiltered sequences, and corresponding weight and Hodge filtrations (following the lines of \cite{Freitag}). In the second part, Sect.~3, we will deal with the Freed-Witten global anomaly condition.
It is known that branes with the Freed-Witten anomalies cannot carry K-theory charges and they are inconsistent. That is the reason why we analyze the Freed-Witten condition for our class of solutions obtained in the first part of the paper and this constitutes a link between its two parts. We will argue, as a result, that the torsion part of the cuspidal cohomology is involved in the global anomaly condition. In fact, in the presence of the cuspidal part all classes of cohomologies can be mapped to a boundary of the space and the cuspidal contribution can then be involved in the global anomaly condition.

\section{The class of general factorized solutions}

For convenience, we recall here some general
results for rotating branes in arbitrary dimensions, supported by a single form charge. In fact these results are all obtained by diagonally oxidizing the rotating black holes constructed in \cite{Cvetic96}. Single-charge branes in supergravity theories are solutions of the Lagrangian \cite{Cvetic99}:
\begin{equation}
e^{-1}{\mathcal L} = R - (1/2)(\partial\phi)^2 - (1/2j!)\, e^{a\phi}
F^2_{j}\,,
\label{Lagrangian}
\end{equation}
where $F_{j}=dA_{(j-1)}$ and $a^2 = 4 -2(j-1)(D-j-1)/(D-2)$.
The Lagrangian (\ref{Lagrangian}) admits an electric $(d-1)$-brane with $d=j-1$ or a magnetic $(d-1)$-brane with $d=D-j-1$. Following \cite{Cvetic99}, we shall consider  here the electric solution only (the magnetic one can be viewed as an
electric solution of the dual $(D-j)$-form field strength
$F_{(D-j)}$). The rotating $p$-brane can be dimensionally reduced on its world-volume spatial coordinates, to give rise to single-charge rotating black holes \cite{Cvetic96}. On the other hand, there is a straightforward procedure  to only oxidize the rotating black hole solutions to give the rotating $p$-branes in higher dimensions. This approach to obtain general single-charge rotating $p$-branes has been used in \cite{Cvetic96}.
According to the procedure for the sphere reduction of generic rotating $p$-branes, we can take the limit of large $p$-brane charge. It then follows that the $(d+1)$-dimensional metric becomes (we use the same notations as in \cite{Cvetic96})
\begin{equation}
ds_{d+1}^2 =-(h_1\cdots h_N)^{-\frac{d-2}{d-1}}\, f\, dt^2 +
(h_1\cdots h_N)^{\frac{1}{d-1}}\Big(
e^{-(D-2)\, \phi}\, f^{-1} \, d\rho^2 +
\rho^2\, d{\bf x}\cdot d {\bf x}\Big)\ .
\label{metric}
\end{equation}
The Einstein-frame metric is given by $ds_E^2 = \exp\{-(D-2)\phi/(d-1)\}\, ds_{d+1}^2$.
This is the metric of an $N$-charge black hole in a domain-wall background.
In the case when $a=0$, the domain wall specializes to AdS$_{d+1}$.

\subsection{Anz\"atze and $2n$-fold space class}
From now on we will consider the class of general factorized solutions and the general construction of the metric (\ref{metric}) assuming that, locally, it is the product of a set of two-dimensional K\"ahler--Einstein metrics
$
ds^2_{n} =\sum ds^2(KE_2)\,,
$
where $d=2n+1$,\, $ds^2(KE_2)$ is a two-dimensional K\"ahler-Einstein metric locally proportional to the standard metric on $S^2$, $H^2$ or $T^2$.\footnote{In the corresponding type IIB solutions {\rm (}$n=2${\rm )} and $D=11$ solutions {\rm (}$n=3${\rm )} one finds that the Killing spinors are independent of the coordinates on $KE_2$. The spaces $KE_2$ can be globally taken, for example, to be $H^2$ or some quotient, $H^2/\Gamma$, the last giving a compact Riemann surface with genus bigger than one, while still preserving supersymmetry {\rm \cite{Waldram2}}.
}
The Ricci form of $ds^2_{n}$ is given by
$
{\cR}= -F
$
where $F$ is the K\"ahler form of the $ds^2(KE_2)$
metric. We assume that, globally, $ds^2_{n}$ extends to the
metric on a space $X_{n}$ which is simply a product of
two-dimensional K\"ahler--Einstein spaces $X_{2n}=\underbrace{KE_2\times\dots\times KE_2}_n$.

In the case $n=1$, $D=4$, $N=8$, one has the $SO(8)$ supergravity solutions arising from $D=11$ supergravity on $S^{7}$ whose black hole solutions are discussed in \cite{Duff99,Cvetic99}.\footnote{Truncations of the gauged supergravities (which include only gauge fields in the Cartan subalgebras) admit the four-charge AdS$_{4}$ black hole solutions \cite{Cvetic99}.
}
The explicit Kaluza-Klein reduction ans\"atze
allow to investigate the embedding of the AdS black
holes of $D=4$ in the respective higher-dimensional
supergravities. It has been shown that a four-dimensional BPS state (whose AdS energy is equal to its electric charge) admits the eleven-dimensional interpretation of a two-brane \cite{Bergshoeff,Duff91} that is rotating in the extra
dimensions. Moreover, the electric charge is equal to the spin in this case.

We should remark that infinite classes of eleven-dimensional supergravity
warped products of AdS$_3$ with an eight-dimensional manifold $X_8$, which are dual to conformal field theories with $N=(0,2)$ supersymmetry (by the AdS-CFT correspondence), have been found in \cite{Waldram1,Waldram2}.
These new solutions are all $S^2$ bundles over
six-dimensional base spaces ${\cB}_6$. A two-sphere bundle can be obtained from the canonical line-bundle over $\cB_6$ by adding a point at infinity to each of the fibers (for details, see \cite{Waldram1}). Spaces $\cB_6$ are products of
K\"ahler-Einstein spaces with various possibilities for the signs of the curvature: the class of the polynomial solution is $\cB_6=KE_2^{(1)}\times KE_2^{(2)}\times KE_2^{(3)}$.

\subsection{Hilbert modular groups and varieties}
\label{Hilbert0}
In this section we consider some examples of general factorized solutions. This class includes the product of $2n$-fold two-space forms which are associated with discrete subgroups of $SL(2, {\mathbb R})^n$ with compact quotient ${\bf H}^n/\Gamma$, the Hilbert modular groups and varieties. We will mostly follow \cite{Freitag} in reproducing the necessary results. Central topics to be considered are the following:
\begin{itemize}
\item{} A discrete subgroup $\Gamma\subset SL(2, {\mathbb R})^n$ with compact quotient ${\bf H}^n/\Gamma$ and the Hilbert modular group. (The corresponding singular cohomology groups $H^\bullet (\Gamma; {\mathbb C})$ have been determined in \cite{Matsushima}.)
\item{} The Eilenberg-MacLane cohomology groups
$ H^{\bullet}(\Gamma_{\bf k}; {\mathbb C})$,
where $\Gamma_{\bf k}$ acts trivially on $\mathbb C$ and the totally real number field ${\bf k}\supset {\mathbb Q}$. The corresponding spaces
are the {\it Hilbert modular varieties}:
${\bf H}^n/\Gamma_{\bf k}   \equiv (\underbrace{
{H}^2\times \ldots \times{H}^2}_n)/\Gamma_{\bf k}$.
\item{} The mixed Hodge structure in the sense of Deligne \cite{Deligne}.
\end{itemize}
The Hilbert modular group, $\Gamma_{\bf k}=SO(2, {\bf o})$, and the corresponding spaces (Hilbert modular varieties) and functions ({\it Hilbert modular forms}) have been actively studied in mathematics (for reference see \cite{Freitag}).
$\Gamma_{\bf k}$ is the group of all $2\times 2$ matrices of determinant 1 with coefficients in the ring $\bf o$ of integers of a totally real number field.
The Eilenberg-MacLane cohomology groups
$H^\bullet(\Gamma_{\bf k}; {\mathbb C})$ are isomorphic to the singular cohomology group of the Hilbert modular variety
$
H^\bullet (\Gamma_{\bf k}; {\mathbb C}) =
H^\bullet ({\bf H}^n/\Gamma_{\bf k};
{\mathbb C}),
$
where, as before, ${\bf H}^n$ denotes the product of $n$ upper half-planes $H^2$ equipped with the natural action of $\Gamma_{\bf k}$. The Hilbert modular variety carries a natural structure as a {\it quasiprojective variety} and its cohomology groups inherit a {\it Hodge structure}. In fact the Hilbert modular group is a simplified example of the cohomology theory of arithmetic groups
and it is the only special case in which the cohomology can be determined explicitly.
\\

\noindent
{\bf Example: Flux quantization for the solution AdS$_3\times X_8$.}
As an example, let us consider the particular solution AdS$_3\times X_8$. Assume that the K\"ahler metric $ds^2(X_4)$ and the four-form $G_4$ are given by
\begin{eqnarray}
ds^2(X_8) & = & \sum_{i=1}^3 \sigma_i ds^2(KE_2)+ \sigma_4 dr^2 + \sigma_5 (D\psi)^2 \,,
\nonumber \\
G_4 & = & g_1F\wedge dr\wedge D\psi +g_2 dr\wedge {\rm Vol}({\rm AdS}_3)\,.
\end{eqnarray}
Here $F= {\rm Vol}(KE_2)$ is the corresponding K\"ahler space form on $KE_2$. These anz\"atze depend on $\sigma_i,\, g_j$, which are functions of $r$ only. The isometry group is
$SO(2,2)\times U(1)$, the first factor corresponds to the symmetries of AdS$_3$ and the latter to shifts of the fiber coordinate $\psi$. Also,
$
D\psi=d\psi+P\,.
$
The twisting of the fibre with coordinate $\psi$
is associated to the canonical $U(1)$ bundle over the six-dimensional base space ${\cB}_6$ given by
$KE_2\times KE_2\times KE_2$.
For the solution AdS$_3\times X_8$ the relevant four-cycles lie in $X_8$. The quantization condition takes the form
\begin{equation}
(2\pi)^{-3}[G_4]
- (1/4)p_1(X_8) \in H^4(X_8,{\mathbb Z})\,,
\end{equation}
where we took the eleven-dimensional Planck length and the $AdS_3$ radius to be one, $p_1(X_8)$ is the first Pontryagin class of $X_4$, and $[G_4]$ denotes the cohomology class of $G_4$ on
$X_8$.\footnote{Recall that the rational cohomology of the direct product of two manifolds, $X$ and $Y$ of dimension $m$ and $n$ respectively, can be calculated by using the K\"unneth formula:
$
H^k(X\times Y; {\mathbb Q}) =\bigoplus_{p+q = k}
(H^p(X; {\mathbb Q})\oplus H^q(Y; {\mathbb Q}))\,.
$
Thus for Betti numbers we have $b^k(X\times Y)= \sum_{p+q=k}b^p(X)b^q(Y)$, and the Euler-Poincar\'{e} characteristic becomes
$
\chi(X\times Y)  =  \sum_{k=0}^{m+n} (-1)^k
[\sum_{p+q=k}b^p(X)b^q(Y)]
= \sum_{p=0}^{m}(-1)^pb^p(X)[\sum_{q=0}^n(-1)^qb^q(Y)]
= \chi(X)\chi(Y)\,.
$
For the product $KE_2^{(1)}\times KE_2^{(2)}$ one gets
$\chi(KE_2^{(1)}\times KE_2^{(2)})=\chi(KE_2^{(1)})
\chi(KE_2^{(2)})$.
}
For the construction of compact solutions we require, as shown in \cite{Waldram1}, $r,\psi$ to parameterize a two-sphere fibred over a compact base $\cB_6 =
{\bf H}^3/\Gamma$.
Besides, if the new solutions have the form of a product of Riemann surfaces, then we get
\begin{equation}
\label{diagram}
   \begin{CD}
      S^2_{r,\psi} @>>>{X}_8 \\
       && @V{\pi}VV \\
       &&& \!\!\!\!\!\!\!\!\!\!{\bf H}^3/\Gamma
\end{CD}
\end{equation}
Taking into account the fibration (\ref{diagram}) one can find an expression for $p_1(X_8)$. We omit this calculation  since the reader can found a similar one in \cite{Waldram1}.
It is clear that the expression for $p_1(X_8)$ depends on alternating sum of the Hodge numbers or the Euler characteristic,
which are listed in Table \ref{Table1} (see Sect.~\ref{Global} for notations).
\begin{table}
\label{Table1}
\begin{center}
\begin{tabular}
{llll}
Table \ref{Table1}. 
\\
\\
\hline
\\
Hilbert modular groups &
Betti and Hodge numbers\\
and Hilbert varieties & 
\\
\\
\hline
\\
A discrete subgroup $\Gamma \subset SL(2, {\mathbb R})^n$ &
$\sum_{m=0}^{2n}\sum_{p+q=m}(-1)^mh^{p, q} = (-2)^n {\rm Vol}({\bf H}^n/\Gamma)$\,, \\
with compact quotient ${\bf H}^n/\Gamma$  & $b^m = C^n_{m/2}$ if $m$ is even, \,
$b^m =0$ if $m$ is odd,
\\
& $b^m = 2^n\cdot {\rm dim}[\Gamma, (2,...,2)] + U(n)$ if $m=n$,
\\
{} & $U(n) = C^n_{n/2}$ ($U(n)=0$) if $n$ is even  (odd)
\\
\\
Arbitrary congruence subgroups of & $\sum_{m=0}^{2n}(-1)^m(b^m_{\rm univ} +
b^m_{\rm Eis} + b^m_{\rm cusp}$),\,\,\, $b^0=b^{2n} = 0$, \\
$SL(2, {\mathbb R})^n$ with quotient $({\bf H}^n)^*/\Gamma$ &
$b^m_{\rm univ}= C^n_{m/2}$ if $m$ is even,
\\
& $b^m_{\rm univ}= 0$\,\,
if\,\, $m$\,\, is \,\, odd,
\\
{}  & $b^m_{\rm Eis}= 0$\,\, if \,\,$0< m < n$,
\\
{} & $b^m_{\rm Eis}= h-1$ \,\,if \,\, $m=2n-1$\,,
\\
{} &
$b^m_{\rm Eis}= h\cdot C^{n-1}_{m-n}$ \,\,if\,\, $n\leq m < 2n-1$,
\\
{} & $b^m_{\rm cusp}= 0$ \,\, if \,\, $m \neq n$,
\\
{} & $b^m_{\rm cusp}= \sum_{p+q=m}h^{p,q}_{\rm cusp}$ \,\,if \,\, $m = n$,
\\
{} & $h^{p,q}_{\rm cusp}=
\sum_{\atop{\scriptstyle b\subset \{1,..., n\}
\atop{\scriptstyle \#b = p}}}
{\rm dim}\,[\Gamma^b, (2,...,2)]_0$
\\
\\
Hilbert modular varieties. & $\sum_{m=0}^{2n}\sum_{p+q=m}(-1)^mh_m^{p, q}(\Gamma)$\,,
\\
Mixed Hodge structure &
$h_m^{p, q} =
h_{m({\rm univ})}^{p, q}+ h_{m({\rm Eis})}^{p, q}+ h_{m({\rm cusp})}^{p, q}$\,, \\
by E. Freitag and C. Ziegler \cite{Freitag}  &
$h^{0,0}_0 = 1$\,, \,\, $h^{p,q}_{m} = 0$\,\, if
\,\, $m= 2n$\,,
\\
& $h_{m({\rm univ})}^{\frac{m}{2}, \frac{m}{2}} =
C^n_{m/2}$,\,\, $h_{m({\rm univ})}^{p,q} = 0$ \,\, otherwise,
\\
{} & $h_{m({\rm Eis})}^{n,n} = h\cdot C^{n-1}_{m-n}$\,\, if\,\, $n<m<2n-1$, \\
{} & $h_{m({\rm Eis})}^{n,n} = h-1$ if $m=2n-1$,
$h_{m({\rm Eis})}^{p,q} = 0$ otherwise,
\\
{} &
$h_{n({\rm cusp})}^{p,q} = h_{\rm cusp}^{p,q} = \sum\!\!\!\!\!\!_{\atop{\scriptstyle b\subset \{1,..., n\}
\atop{\scriptstyle p=n-\# b, q=\# b}}}\!\!\!{\rm dim}\,[\Gamma^b, (2,..., 2)]_0$
\\
{} & if\,\, $p+q =n$,\,\, $h^{p,q}_{m({\rm cusp})} = 0$ \,otherwise
\\
\\
\hline
\end{tabular}
\end{center}
\end{table}
\\

\noindent
{\bf Duality relations.}
The class $\cB_{2n}={\bf H}^n/\Gamma$ is particular interesting because it admits the duality relations in type II string theory. Indeed, let us
turn now to spaces $X_{2n+k} = \cB_{2n}\times T^k$. For spaces ${T}^k$ one has
$
H^p({T}^k, {\mathbb Z}) =
H_p({T}^k, {\mathbb Z})= {\mathbb Z}^{C^k_p}
$,
where
$
{C}^k_p \equiv p!/k!(p-k)!.
$
Because of the K\"unneth formula we get $K_{j}(T^k) \cong K_{j}(S^1)^{\oplus 2^{k-1}}\cong
{\mathbb Z}^{\oplus 2^{k-1}}$ for $j =0,1$. Thus the following isomorphisms valid:
\begin{eqnarray}
K_0({\cB}_{2n} \times {T}^k)  & \cong &
\left(\widetilde{K}_0 (\cB_{2n}) \oplus
K_{1}(\cB_{2n}) \oplus {\mathbb Z}\right)^{\oplus 2^{k-1}}\,,
\nonumber \\
K_1 (\cB_{2n} \times {T}^k) & \cong &
\left(K_{1}(\cB_{2n}) \oplus
\widetilde{K}_0 (\cB_{2n}) \oplus {\mathbb Z}\right)^{\oplus 2^{k-1}}
\Longrightarrow
K_0 (\cB_{2n} \times {T}^k)
\nonumber \\
& \cong &
K_1 (\cB_{2n} \times {T}^k) \ .
\label{Isom}
\end{eqnarray}
Here $\widetilde{K}_j(X)$ is the reduced topological K-groups of $X$.
\footnote{
Taking into account that a vector bundle
over a point is just a vector space,
$K({\rm pt})= {\mathbb Z}$,
we can introduce a reduced K-groups in which the topological space consisting of a single point has trivial cohomology,
${\widetilde{K}}({\rm pt})=0$, and also
${\widetilde {K}}(X)=0$
for any contractible space $X$. Let us consider the collapsing and inclusion maps:
$ p: X\rightarrow {\rm pt}\,,\,\,\iota: {\rm
pt}\hookrightarrow X $ for a fixed base point of $X$.
These maps
induce an epimorphism and a monomorphism of the corresponding
K-groups: $ p^*: K({\rm pt}) = {\mathbb Z}\rightarrow K(X)\,,$ $\iota^*: K(X)\rightarrow K({\rm pt})= {\mathbb Z}.$ The exact sequences of groups are:
$
0\rightarrow {\mathbb Z} \stackrel{p^*}{\rightarrow} K(X)
\rightarrow {\widetilde{K}}(X)
\rightarrow 0\,,\,
0 \rightarrow {\widetilde {K}}(X)\rightarrow K(X)
\stackrel{\iota^*}{\rightarrow} {\mathbb Z}\,.
$
The kernel of the map $i^*$ is called the {\it reduced K-theory group} and is denoted by
${\widetilde {K}}(X)$, ${\widetilde {{K}}}(X) = {\rm ker} \iota^* = {\rm coker}\,p^*$. There is the fundamental
decomposition $K(X)= {\mathbb Z}\oplus \widetilde{{K}}(X)$.
When $X$ is not compact, we can define $K^c(X)$, the K-theory with compact support; it is isomorphic to ${\widetilde{{K}}}(X)$.
}
The isomorphism (\ref{Isom}) is in fact the T-duality and describes a relationship between Type IIB and Type IIA D-branes on the space $\cB_{2n}\times {T}^k$. This isomorphism exchanges wrapped D-branes
with unwrapped D-branes. In adition the powers of
$2^{k-1}$ give the expected multiplicity of D-brane charges arising from wrapping all higher stable branes on various cycles of the torus ${T}^k$.

\section{The global anomaly condition}
\label{Global}

This section is motivated by the fact that branes with the Freed-Witten anomalies cannot carry K-theory charges and they are inconsistent. That is the reason why we analyze here the Freed-Witten condition for our class of solutions obtained in the previous section.
Let us consider a bundle whose fiber $\xi$ is $(p-1)$-connected.
This means that, for $k<p$, the $k$-th homotopy group
$\pi_{k< p}(\xi)$ of $\xi$ vanishes. Let the $p$th homotopy group $\cG$ be non-trivial and $\pi_{k=p}(\xi) = \cG$. Then the $\xi$ bundle can be specified by a degree $(p+1)$ characteristic class in the cohomology with coefficients in $\cG$, $\omega_{p+1}\in H^{p+1}(X;\cG)$, and some characteristic classes of higher degree. If all of the homotopy classes of $\xi$ of degree higher than $p$ vanish, then the bundle is characterized by $\omega_{p+1}$.
If $\cG = {\mathbb Z}_2$ and $p=0$, then a non-vanishing characteristic class $\omega_1\in H^1(X; {\mathbb Z}_2)$ is associated with the existence of a so-called {\it spin} structure on $X$. For the circle bundle, $p=1,\, \cG= {\mathbb Z}$, we have a single characteristic class $c_1=\omega_2\in H^2(X; {\mathbb Z})$ which is called {\it the first Chern class}.
For every degree $(p+1)$ characteristic class there is a degree $(p+2)$ obstruction to a lift. The obstruction to the existence of a spin structure is the third Stiefel-Whitney class $\omega_3\in H^3(X; {\mathbb Z})$. The class $\omega_3$ is always ${\mathbb Z}_2$ torsion. Unlike $\omega_2$, the class $\omega_3$ always has a lift to cohomology with integral coefficients, which is denoted by $W_3\in H^3(X; {\mathbb Z})$, and also has always ${\mathbb Z}_2$ torsion. $W_3$ is defined to be the Bockstein homomorphism $\beta$ of $\omega_2$, $W_3= \beta\omega_2$. If the third Stiefel-Whitney class of the tangent bundle $TX$ is equal to zero, then $X$ is said to be a spin manifold (and a spin lift of the tangent bundle exists).

A free brane can wrap any homologically nontrivial cycle in $X$.
It is known that a brane wrapping a representable cycle ({\it spin$^C$cycle}) carries a K-theory charge if and only if its Freed-Witten anomaly vanishes. Nevertheless, some K-theory charges
are only carried by branes that wrap non-representable cycles.
The Freed-Witten anomaly condition is
\begin{equation}
W_3(X) + [H]\vert_{X} =0 \,\,\,\, {\rm in }
\,\,\,\, H^3(X;\mathbb Z)\,.
\label{condition}
\end{equation}
Here $H$ is the pullback of the NSNS three-form to the space worldvolume $X$.\footnote{Note that in the de Rham theory
$[H]_{DR}\vert_{X}=0$. In the bosonic string case we must impose condition (\ref{condition}) without the $W_3(X)$ term.
}
The Freed-Witten anomaly is a necessary but not sufficient condition for the homology class of a D-brane to lift to a twisted K-theory.
\footnote{
When the NS 3-form $H$ is topologically trivial, the Freed-Witten anomaly is a necessary and sufficient condition for all D$p$-branes except for D6-branes \cite{Evslin}.
}
To provide an answer to the question if D-branes can wrap non-representable cycles, one must look at the worldsheet theory of fundamental strings, impose boundary conditions corresponding to a singular representative of the cycle (the space $\cB_{2n}$ in our case) and then check for inconsistencies, such as a failure of BRST invariance. The complete analysis is quite complicated and we shall leave it for a forthcoming paper. Here we will start with some considerations on the global anomaly condition, which constitute an unavoidable part of it.
\\

\noindent
{\bf Circle bundles on Riemannian surfaces.}
A D-brane wrapping homologically nontrivial cycle $Y$ can nevertheless be unstable, if for some $Y'\subset X$
the following condition holds \cite{Maldacena}:
\begin{equation}
PD(Y \subset Y') = W_3(Y') + [H]\vert_{Y'}\,.
\label{COND}
\end{equation}
In Eq. (\ref{COND}) the left hand side denotes the Poincar\'{e} dual of $Y$ in $Y'$. (In bosonic strings the question of stability is more complicated because they always include tachyons.) One can use a mathematical algorithm known as the Atiyah-Hirzebruch spectral sequence to determine which homology classes lift to K-theory classes, that is, to determine which D-branes are unstable and which are not allowed.

First we consider the simple case of a Riemann surface $\Sigma_g=H^2/{\Gamma}_g$ of genus $g$.
For a Riemann surface of genus $g$,  $H^2({\Sigma}_g, {\mathbb Z})= {\mathbb Z}$ and topologically circle bundles are classified by an integer $j$. The cohomology of the total space $X_3$ is given by
\begin{enumerate}[{-}]
\item{}
$j=0$ (trivial line bundle):
$H^0(X_3,{\mathbb Z})= {\mathbb Z}$,\,\,
$H^1(X_3,{\mathbb Z})={\mathbb Z}^{2g+1}$,\,\,
$H^2(X_3,{\mathbb Z})={\mathbb Z}^{2g+1}$,\,\,
$H^3(X_3,{\mathbb Z})={\mathbb Z}$;
\item{}
$j\neq 0$ (the Chern class equal to $j$):
$H^0({X_3},{\mathbb Z})={\mathbb Z}$,\,\,
$H^1(X_3,{\mathbb Z})={\mathbb Z}^{2g}$,\,\,
$H^2(X_3,{\mathbb Z})={\mathbb Z}^{2g}\oplus
{\mathbb Z}_j$,\,\,
$H^3(X_3,{\mathbb Z})={\mathbb Z}$\,.
\end{enumerate}
Then, the untwisted ($H=0$) and the twisted ($H=k$) K-groups are \cite{Bouwknegt1,Bytsenko}:
\begin{eqnarray}
K^0(X_3,{H=0})= H^0(X_3,{\mathbb Z})\oplus H^2(X_3,{\mathbb Z}) & = &
\, \left\{
\begin{matrix} {\mathbb Z}^{2g+2}\,\,\,\,\,&\textup{if $j=0$}\,,\cr
{\mathbb Z}^{2g+1}\oplus{\mathbb Z}_j\,\,\,\,\, &\textup{if $j\neq 0$}\,,
\end{matrix}
\right.\nonumber\\
K^1(X_3,{H=0})= H^1(X_3,{\mathbb Z})\oplus H^3(X_3,{\mathbb Z}) & = &
\, \left\{
\begin{matrix} {\mathbb Z}^{2g+2}\,\,\,\,\,\,\,\,\,\,\,\,\,\,\,\,\,\,\,
&\textup{if $j=0$}\,,\cr
{\mathbb Z}^{2g+1}\,\,\,\,\,\,\,\,\,\,\,\,\,\,\,\,\,\,\,
&\textup{if $j\neq 0$}\,.
\end{matrix}
\right. \nonumber \\
K^0(X_3,{H=k}) = H^2(X_3,{\mathbb Z}) & = &
\left\{
\begin{matrix} {\mathbb Z}^{2g+1}\,\,\,\,\,\,\,\,\,\,\,\, &\textup{if $j=0$}\,,\cr
{\mathbb Z}^{2g}\oplus{\mathbb Z}_j\,\,\,\,\,\,\,\,\,\,\,\,&\textup{if $j\neq 0$}\,,
\end{matrix}
\right.\nonumber \\
\!\!\!\!\!\!\!\!\!\!\!\!\!\!\!\!\!\!
K^1(X_3,{H=k})= H^1(X_3,{\mathbb Z})\oplus H^3(X_3,{\mathbb Z})/kH^3(X_3,{\mathbb Z}) & = &
\left\{
\begin{matrix} {\mathbb Z}^{2g+1}\oplus{\mathbb Z}_k\,\,\,\,\, &\textup{if $j=0$}\,,\cr
{\mathbb Z}^{2g}\oplus{\mathbb Z}_k\,\,\,\,\, &\textup{if $j\neq 0$}\,.
\end{matrix}
\right.
\label{K-gr}
\end{eqnarray}
In fact T-duality is the interchange of $j$ and $k$. Results in the twisted K-groups $K^0(X_3,H)$ and $K^1(X_3,H)$ are interchanged, which corresponds to the
fact that RR field strengths are classified by $K^0(X_3,H)$ in type IIA string theory and by $K^1(X_3,H)$ in the IIB one (compare with the results in (\ref{Isom})).
This means that, applying the isomorphism between the two K-groups, one can find the new RR field strengths from the old ones. Indeed, one simply interchanges the ${\mathbb Z}^{2g}$ between $H^1$ and $H^2$ and the rest of the cohomology groups are swapped $H^0\leftrightarrow H^1,\ H^2\leftrightarrow H^3$.

Let us now analyse twisted K-groups of circle bundles over two-manifolds, the K-groups being determined by the Atiyah-Hirzebruch spectral sequence. The differential $d_3$ in K-theory (the reader can find some homological and K-theory methods applied to hyperbolic cycles in \cite{Bonora,Bytsenko}) has the form \cite{Atiyah,Rosenberg}
\begin{equation}
d_3 = Sq^3,\,\,\,\,\,\, d_3 = Sq^3 + [H]\,\,\,\,\,\,{\rm for\,\,\, twisted\,\,\, K\!-\!theory}\,,
\end{equation}
where the Steenrod square $Sq^3$ (for the cohomology operations see the Appendix) takes an integral class in the $k$th cohomology to a class in the $(k+3)$rd cohomology, as does the cup product with $[H]$. A necessary (but not sufficient) condition for the vanishing of $(W_3(X) + [H]|_X)$ on a worldvolume is that the flux has to be in the kernel of the spectral differential sequence
$
d_3 = Sq^3 + [H] \cup.
$
In the de Rham theory, for example, we obtain the simple expression
$
d_3(\omega) = [H]\wedge \omega\,.
$
Unlike the cup product with $[H]$, $Sq^3$ is only nontrivial when acting on ${\mathbb Z}_2$ torsion components of $H^k(X)$ and the image is, likewise,  a ${\mathbb Z}_2$ torsion component of $H^{k+3}(X)$ always. Generally speaking, in passing from $K(X)$ to the full K-theory group one needs to solve an extension problem to obtain the correct torsion subgroup. This has interesting physical applications\footnote{In the physical context, the extension problem can be quite important in the presence of orbifolds (see \cite{Bergman} for a discussion of this point).}

In our case it is convenient to consider the first differential $d_3=Sq^3+H$ of the sequence only. If $H^{\rm even}(X,{\mathbb Z})$ and $H^{\rm odd}(X,{\mathbb Z})$ are the even and odd
cohomology classes of the manifold $X$, then the twisted K-groups are
\begin{equation}
K^0(X,H)=\frac{\textup{Ker}(H\cup:H^{\rm even}\longrightarrow H^{\rm odd})}
{H\cup H^{\rm odd}(X,{\mathbb Z})}\,,\quad
K^1(X,H)=\frac{\textup{Ker}(H\cup:H^{\rm odd}\longrightarrow H^{\rm even})}
{H\cup H^{\rm even}(X,{\mathbb Z})}\,.
\end{equation}
We can specialize to the case when
$\Sigma_g \equiv G/K,\, G= {\mathbb R}^2$,\, $K = \{e\}$ and $g=1$, with $\Gamma_1$ being ${\mathbb Z}^2$. The graded groups are given by \cite{Bytsenko}
\begin{equation}
Gr(K^0(\Sigma_g)) =
\bigoplus_j E^{2j}_{\infty}(\Sigma_g)\,,\,\,\,\,\,\,\,\,\,\,\,\,
Gr(K^1(\Sigma_g)) =
\bigoplus_j E^{2j+1}_{\infty}(\Sigma_g)\,.
\end{equation}
In two dimensions the Chern character is an isomorphism over the integers and, therefore, we can determine the cohomology class lift to the K-theory class. In particular,
$
K^0(\Sigma_g)\cong H^0(\Sigma_g, {\mathbb Z})\oplus
H^2(\Sigma_g, {\mathbb Z})\cong {\mathbb Z}^2\,,
\,
K^1(\Sigma_g)\cong H^1(\Sigma_g, {\mathbb Z})\cong
{\mathbb Z}^{2g}\,.
$
\\

\noindent
{\bf Hilbert modular varieties.}
A D-brane that wraps a nontrivial cycle carries a charge that corresponds to the homology class of the cycle. In order to compute the partition function it is sufficient to restrict one's attention to equivalence classes of anomaly free branes. Thus D-branes can be classified by a quotient of a subset of homology. It has been argued \cite{Maldacena} that this quotient of a subset is precisely twisted K-theory. Motivated by anomalies we shall now analyze the Freed-Witten anomaly condition for the class of solutions ${\bf H}^n/\Gamma$.

Recall some results on the Hodge (F) and weight (W) filtrations on $H^m(\Gamma)$ (the reader can consult the monography \cite{Freitag} for the Hilbert modular forms). The Deligne's theory contains the classical Hodge theory. Indeed if $X=\overline{X}$ is compact, then the mixed Hodge structure on $H^m(X, {\mathbb C})$ (singular cohomology) is equivalent to the Hodge decomposition
$
H^m(X; {\mathbb C})=\bigoplus_{p+q=m}\cH^{p,q}(X)\,.
$
If $p+q\neq m$ the $h^{p,q}_m(X)$ are zero and coincide with the Hodge numbers $h^{p,q}:={\rm dim}_{\mathbb Z}\cH^{p,q}(X)$ in the remaining case.

Let $\Gamma$ act freely on ${\bf H}^n$, then one can show that the Hodge filtration $F=\emptyset$ and
$X_{2n}={\bf H}^n/\Gamma$. If $m> 0$ then the cohomology $H^m({\bf H}^n/\Gamma, {\mathbb C})$ has the decomposition
\begin{equation}
H^m({\bf H}^n/\Gamma; {\mathbb C}) = H^m(\Gamma)
= H^m_{\rm univ}(\Gamma)\oplus H^m_{\rm cusp}(\Gamma)
\oplus H^m_{\rm Eis}(\Gamma)\,.
\end{equation}
Here each class of $H^m(\Gamma)$ is written as a sum of the Eisenstein (Eis) cohomology class, the class of square integrable differential forms\footnote{Square integrable differential forms can always be presented by square harmonic forms.
}
which can be further decompose as universal (univ) and cuspidal (cusp) parts (see for example \cite{Freitag}).
The Hodge and weight filtration on $H^m(\Gamma)$ can be defined as filtrations induced by $({L}^{\bullet}, F, W)$ on
$H^m(\Gamma {L}^{\bullet})\stackrel{\sim}{\hookrightarrow} H^m(\Gamma)$,
\begin{equation}
W^{\ell} H^m(\Gamma {L}^{\bullet}):= \frac{W^{\ell} {L}^m \cap Z^m}
{W^{\ell} {L}^m \cap B^m}\,,\,\,\,\,\,\,\,\,\,\,\,
F^{p} H^m(\Gamma {L}^{\bullet}) := \frac{F^{p} {L}^m\cap Z^m}
{F^{p}{L}^m\cap B^m}\,,
\end{equation}
where we refer to ${L}^{\bullet}$ as the differentiable logarithmic de Rham complex \cite{Deligne}, and
$
Z^m:= {\rm Ker}({L}^{m}\stackrel{d}{\longrightarrow}
{L}^{m+1})\,,
$
$
B^m:= {\rm Im}({L}^{m-1}\stackrel{d}{\longrightarrow}
{L}^{m})\,.
$
Finally the Hodge filtration on $H^m_{\rm squ}(\Gamma)$ is
\begin{eqnarray}
F^pH^m_{\rm squ}(\Gamma) & = & F^pH^m_{\rm univ}(\Gamma)
\oplus F^pH^m_{\rm cusp}(\Gamma)\,,
\nonumber \\
F^pH^m_{\rm univ}(\Gamma) & = &
H^m_{\rm univ}(\Gamma)\,\,\,\,(\{ 0 \})\,\,\,\, {\rm for}\,\,\, p\leq m/2\,\,\,\, (p > m/2)\,,
\nonumber \\
F^pH^m_{\rm cusp}(\Gamma) & = &
\bigoplus_{p'\geq p}\bigoplus_{\atop{\scriptstyle \,\, b\subset \{1,..., n\}
\atop{\scriptstyle \,\, p'= n- \#b}}}
[\Gamma^b, (2,...,2)]_0\,\,\,\,(\{ 0 \})\,\,\,\,
{\rm for}\,\,\, m=n \,\,\, (m\neq n)\,.
\end{eqnarray}
The main results are given in Table \ref{Table00}.
\begin{table}\label{Table00}
\begin{center}
\begin{tabular}
{llll}
Table \ref{Table00}. Hodge decomposition
$H^m(\Gamma)=\bigoplus_{p,q} \cH_m^{p, q}(\Gamma)$
\\
\,\,\,\,\,\,\,\,\,\,
\,\,\,\,\,\,\,\,\,\,
of the mixed Hodge structure $(H^m(\Gamma), F, W)$
\\
{} & {} \\
\hline
\\
\\
$\cH_m^{p,q}(\Gamma) \,\,\,\,\,\,\,\,\, =
\cH_{m(\rm univ)}^{p, q}(\Gamma)\oplus
\cH_{m(\rm cusp)}^{p, q}(\Gamma)\oplus
\cH_{m(\rm Eis)}^{p, q}(\Gamma)$
&
$\cH_{0}^{0, 0}(\Gamma) \,\,\,\,\,\,\,\, =  H^0(\Gamma)$ \\
$\cH_{m(\rm univ)}^{\frac{m}{2}, \frac{m}{2}}(\Gamma)  =
H_{\rm univ}^{m}(\Gamma)$
&
$\cH_{m(\rm univ)}^{p, q}(\Gamma)  = \{0\}$
otherwise \\
$\cH_{n(\rm cusp)}^{p, q}(\Gamma)  =
\bigoplus\!\!\!_{\atop{\scriptstyle b\subset \{1,..., n\}
\atop{\scriptstyle p = n- \#b, q= \#b}}}
\![\Gamma^b, (2,...,2)]_0$
&
$\cH_{m(\rm cusp)}^{p, q}(\Gamma)  =  \{0\}$
otherwise \\\
$\cH_{m(\rm Eis)}^{n, n}(\Gamma) =
H_{\rm Eis}^{m}(\Gamma)$
&
$\cH_{m(\rm Eis)}^{p, q}(\Gamma) \,\, =  \{0\}$
\,\, otherwise
\\
\\
\hline
\end{tabular}
\end{center}
\end{table}
In analyzing the global anomaly condition, recall that for any $X$, $m$, and associated abelian group $G$, the following result holds (the universal coefficient theorem):
the homology and cohomology group of $X$ with coefficients in $G$ has a splitting:
\begin{eqnarray}
H_m(X; G) & \cong & H_m(X)\otimes G
\oplus {\rm Tor}\, (H_{m-1}(X; G))\,,
\nonumber \\
H^m(X; G) & \cong & H^m(X)\otimes G
\oplus {\rm Tor}\, (H^{m+1}(X; G))\,,
\nonumber \\
H^m(X; G) & \cong &
{\rm Hom}\,(H_m(X); G)\oplus {\rm Ext}\,
(H_{m-1}(X); G)\,.
\end{eqnarray}
Here $H^m(X)$\,\,($H_m(X))$ are the cohomology (homology) groups with integer coefficients.
The (splitting) isomorphisms given by the universal coefficient theorem are said to be unnatural isomorphisms. The following maps of exact sequences are natural:
\begin{equation}
\begin{array}{ccccccccc}
0 & \longrightarrow & \!\!\!\!\!\!\!\!\!\!\!\!\! H_m(X)\otimes G  & \longrightarrow & H_m(X; G) & \longrightarrow &
\!\!{\rm Tor}(H_{m-1}(X); G)  & \longrightarrow & 0\,
\\
0 & \longrightarrow & \!\!\!\!\!\!\! H^m(X; {\mathbb Z})\otimes G   & \longrightarrow & H^m(X; G)  & \longrightarrow &
\,\,\,\,\,{\rm Tor}(H^{m+1}(X; {\mathbb Z}), G) & \longrightarrow & 0\,
\\
0 & \longleftarrow & {\rm Hom}(H_m(X); G) & \longleftarrow & H^m(X; G)
& \longleftarrow & {\rm Ext}(H_{m+1}(X); G) & \longleftarrow & 0\,
\end{array}
\label{sec}
\end{equation}
For example, the first exact sequence of (\ref{sec}) can be deduced
as follows. Let $G={\mathfrak G}_1/{\mathfrak G}_2$, where ${\mathfrak G}_1$ and ${\mathfrak G}_2$ are the abelian groups. It is clear that
$H_m(X; {\mathfrak G}_j)= H_m(X)\otimes {\mathfrak G}_j$, where the group
${\mathfrak G}_j$ is a sum of groups $\mathbb Z$. Thus,
$
H_m(X; {\mathfrak G}_j) = H_m(X; {\mathbb Z}\oplus {\mathbb Z}\oplus
\cdots ) = H_m(X)\oplus H_m(X)\oplus \cdots = H_m(X)\otimes
{\mathfrak G}_j.
$
Let us consider the fifth-term fragment of a sequence
\begin{eqnarray}
&&
\!\!\!\!\!\!\!\!\!\!\!\!\!\!\!\!
H_m(X; {\mathfrak G}_2)\,\,\,\,\longrightarrow
H_m(X; {\mathfrak G}_1)\,\,\,\,\longrightarrow
H_m(X; G)\longrightarrow
H_{m-1}(X; {\mathfrak G}_2)\,\,\,\longrightarrow
H_{m-1}(X; {\mathfrak G}_1) \,\,\,\,\,\, \Longrightarrow
\nonumber \\
&&
\!\!\!\!\!\!\!\!\!\!\!\!\!\!\!\!
H_m(X)\otimes {\mathfrak G}_2\longrightarrow
H_m(X)\otimes {\mathfrak G}_1\longrightarrow
H_m(X; G)\longrightarrow
H_{m-1}(X)\otimes {\mathfrak G}_2\longrightarrow
H_m(X)\otimes {\mathfrak G}_1
\label{sec1}
\end{eqnarray}
Note that any fifth-term exact sequence
$
A\stackrel{\sigma}{\rightarrow} B\rightarrow C\rightarrow D
\stackrel{\tau}{\rightarrow} E
$
can be transformed into a short exact sequence
$
0\rightarrow {\rm Coker}\,\sigma\rightarrow C
\rightarrow {\rm Ker}\,\tau\rightarrow 0,
$
where ${\rm Coker}\,\sigma = B/{\rm Im}\,\sigma.
$
This transformation converts the last fifth-term sequence into the first short sequence in (\ref{sec}). The second sequence of (\ref{sec}) can be similarly derived, while the verification of the last sequence requires more complicated arguments.

Taking into account the Hodge decomposition (see Table \ref{Table00}) note that for a strictly co-compact subgroup $\Gamma \subset SL(2, {\mathbb R})^n$ the cohomology group $H^{m=3}(\Gamma)$ vanishes. From the second sequence of (\ref{sec}) it follows that for $m=3$ and $G= {\mathbb C}$ the group
$H^3(X_{2n}={\bf H}^n/\Gamma; {\mathbb Z})\otimes {\mathbb C}$ vanishes.
Therefore, we conclude that, in the case of a co-compact group $\Gamma$ ,the torsion-free group of cohomology $H^3({\bf H}^n/\Gamma; {\mathbb Z})$ is trivial (e.g., the class of $[H]_{X_{2n}}$ is trivial).
A more complicated situation occurs in presence of the cuspidal part of the total group of cohomology. For a discrete subgroup $\Gamma_\kappa \subset SL(2, {\mathbb R})^n$ and each boundary point $\kappa \in {\mathbb R}\cup\{\infty\}$ there exist parabolic elements in the stabilizer $\Gamma_\kappa$. All the cohomology class can be mapped to a boundary of the space. Therefore the cuspidal contribution
\begin{equation}
H^{n}_{\rm cusp}(\Gamma)=
\bigoplus_{p+q=n}
\cH_{n(\rm cusp)}^{p, q}(\Gamma)
=
\bigoplus_{p+q=n}
\!\bigoplus_{\atop{\scriptstyle b\subset \{1,..., n\}
\atop{\scriptstyle p = n- \#b, q= \#b}}}
\!\!\!\!\!\![\Gamma^b, (2,...,2)]_0 \,\,\,\, {\rm for}\,\,\,\, n=3
\end{equation}
could be involved in the global anomaly condition (\ref{condition}).

Note finally  that, for a rigorous analysis of the global anomaly,
one plainly needs to have an explicit expression for the higher differentials $\{d_\ell\}_{\ell=3}^{2n-1}$. Actually an expression for the differential $d_3$ is available, but not very much is known about the higher differentials, in general. In this paper we have restricted ourselves to the case of the differential $d_3$, only.

\subsection*{Acknowledgements}

The first author would like to acknowledge the ESF Research Network CASIMIR and the Conselho Nacional de Desenvolvimento Cient\'ifico e Tecnol\'ogico (CNPq) for financial support. The second one was partially funded by Ministerio de Educaci\'on y Ciencia, Spain, project FIS2006-02842, and AGAUR (Generalitat de Catalunya), contract 2009SGR-994 and grant 2010BE-100058. EE's research was performed in part while on leave at Department of Physics and Astronomy, Dartmouth College, 6127 Wilder Laboratory, Hanover, NH 03755, USA.

\section{Appendix: Cohomology operations}
\label{Steenrod}
Algebraic methods for the computation of cohomology groups rely on the concept of a {\it cohomology operation}
$
{\mathfrak m}:\, H^q({\mathfrak C}, {\mathfrak D}; G_1)\rightarrow H^r({\mathfrak C}, {\mathfrak D}; G_2)\,.
$
Here ${\mathfrak C}$ is a cell complex and ${\mathfrak D}\subset {\mathfrak C}$ a subcomplex (for example, ${\mathfrak D} = {\mathfrak C}^{\ell-1}$, the $(\ell+1)-$skeleton of
${\mathfrak C}$). The following properties hold \cite{Dubrovin}:
\begin{enumerate}[{-}]
\item{}
The map $\mathfrak m$ should be defined for each pair, $C, D$, of complexes;
\item{}
$\mathfrak m$ should commute with a continuous map
$f:\, ({\mathfrak C}, {\mathfrak D})\rightarrow ({\mathfrak C}', {\mathfrak D}'),  i.e., {\mathfrak m} f =f{\mathfrak m}$\,.
\end{enumerate}
For $k\geq 2$, the maps $d_k : \cap_{\ell<k}\,{\rm Ker}\, d_\ell\rightarrow H^{q+1}/\cup_{\ell< k}\,{\rm Im}\,d_\ell$ are, in general, neither single-valued nor everywhere defined
({\it higher} cohomology). Knowing the structure of the $H^\bullet(-; {\mathbb Z}_p)$ groups and the actions of $d_k$, we can determine the structure of the quotients $H^\bullet(-; {\mathbb Z}_p)/{G}_p$ of the integral relative homology groups, where
${G}_p$ is the subgroup of elements of finite order not divisible by $p$. Therefore, knowledge of the operators $d_k$ acting on the cohomology groups $H^\bullet(-; {\mathbb Z}_p)$, for all $p$ (or the dual operators on the homology groups $H_\bullet(-; {\mathbb Z}_p)$), permits establishing the structure of the integral cohomology (or homology) groups.
Operators $d_k$ possess the following properties:
\begin{enumerate}[{-}]
\item{}
$d_k$ is defined on the $H^q(-; {\mathbb Z}_p)$ (or certain subgroups of these) and are homomorphisms;
\item{}
$H^q(-; {\mathbb Z}_p)\stackrel{d}{\longrightarrow}
H^{q+1}(-; {\mathbb Z}_p)$,\, where $d$ is the homomorphism of exact sequence of the pair $({\mathfrak C}, {\mathfrak D})\,,\,\, d\circ d_k = d_k\circ d$. A cohomology operation with these properties is called {\it stable}.
\end{enumerate}
\begin{theorem} {\rm (}Steenrod{\rm )}\,
Let $p=2$, then for each $k\geq 0$ there is the stable cohomology operation, a {\it Steenrod square}, denoted by $Sq^k$, which is a homomorphism $Sq^k: \, H^q(-; {\mathbb Z}_2) \rightarrow H^{q+k}(-; {\mathbb Z}_2)$\,,
$\forall k$, and which has the following properties:
\begin{eqnarray}
Sq^kx & = & 0\,\,\,\,\,\,\,\,{\rm for}\,\,\,\,\, q<k;
\,\,\,\,\,\,\,\,\,\,\,\,\,\,\,
Sq^0 \equiv 1;
\nonumber \\
Sq^kx & = & x^2\,\,\,\,\,{\rm for}\,\,\,\,\, q=k;
\,\,\,\,\,\,\,\,\,\,\,\,\,\,\,
Sq^1x = d_1x\,;
\nonumber \\
Sq^k(xy) & = & \!\!\!\sum_{j+\ell = k}Sq^j(x) Sq^\ell (y)\,.
\nonumber
\end{eqnarray}
Let $p> 2$, then for all $k\geq 0$ there is the stable cohomology operation
$
St_p^k: \, H^q(-; {\mathbb Z}_p)\rightarrow
H^{q+2k(p-1)}(-; {\mathbb Z}_p)\,
$
such that
\begin{eqnarray}
&&
St_p^kx =0\,\,\,\,\,\,\,\,{\rm for}\,\,\,\,\, q<2k;
\,\,\,\,\,\,\,\,\,\,\,\,\,\,\,
St_p^0 \equiv 1;
\nonumber \\
&&
St_p^kx =x^p\,\,\,\,\,{\rm for}\,\,\,\,\, q=2k;
\,\,\,\,\,\,\,\,\,\,\,\,\,\,\,
St_p^k(xy) = \sum_{j+\ell = k}St_p^j(x) St_p^\ell (y)\,.
\nonumber
\end{eqnarray}
\end{theorem}
The operations $Sq^k\,,\,St_p^k$ are called {\it Steenrod operations}. In mod $k$ cohomology all stable cohomology operations are composite of the Steenrod operations. These facts, together with some others, form the basis of a systematic procedure of computation of the torsion subgroups of homotopy groups.


\begin{thebibliography}{99}

\bibitem{cv1} E. Verlinde, {\it On the holographic principle in a radiation dominated universe}, arXiv:hep-th/0008140v2; R.-G. Cai, {\it Cardy-Verlinde formula and AdS black holes}, Phys. Rev. D {\bf 63} (2001) 124018;
S. Nojiri, S. D. Odintsov and  S. Ogushi, {\it Friedmann-Robertson-Walker brane cosmological equations from the five-dimensional bulk (A)dS black hole}, Int. J. Mod. Phys. A {\bf 17} (2002) 4809.

\bibitem{gpp1}  G. W. Gibbons, M. J. Perry and C. N. Pope, {\it AdS/CFT Casimir energy for rotating black holes}, Phys. Rev. Lett. {\bf 95} (2005) 231601;
Y. S. Myung, {\it Entanglement system, Casimir energy and black hole}, Phys. Lett. B {\bf 636} (2006) 324.

\bibitem{Cvetic99}
M. Cveti\v{c}, M. J. Duff, P. Hoxha, J. T Liu, H. Lu, J. X. Lu, R. Martinez-Acosta, C. N. Pope, H. Sati, and T. A. Tran, {\it Embedding AdS Black Holes in Ten and Eleven Dimensions}, Nucl. Phys. B {\bf 558} (1999) 96; [arXiv:hep-th/9903214].

\bibitem{Bytsenko05}
A. A. Bytsenko, M. E. X. Guimar\~aes and J. A. Helay\"{e}l-Neto,
{\it Hyperbolic Space Forms and Orbifold Compactification in M-Theory}, PoS WC2004 (2004) 017; [arXiv:hep-th/0502031].

\bibitem{Bytsenko09}
A. A. Bytsenko, {\it Global anomaly and a family of structures on fold product of complex two-cycles}, in: ``Geometrical Methods in Physics'', Proceedings of the XXVIII Workshop on Geometrical Methods in Physics, American Institute of Physics, {\bf 1191} (2009) 59; [arXiv:hep-th/0910.5178].

\bibitem{Kim1}
N.~Kim, {\it AdS(3) solutions of IIB supergravity from D3-branes},
JHEP {\bf 0601} (2006) 094; [arXiv:hep-th/0511029].

\bibitem{Kim2}
N.~Kim and J.~D.~Park, {\it Comments on AdS(2) solutions of D = 11 supergravity}, JHEP {\bf 0609} (2006) 041;
[arXiv:hep-th/0607093].

\bibitem{Cvetic1}
M.~Cvetic, M. J. Duff, P. Hoxha, J. T. Liu and H. L\"u,
{\it Embedding AdS black holes in ten and eleven dimensions},
Nucl. Phys. B {\bf 558} (1999) 96; [arXiv:hep-th/9903214].


\bibitem{Freitag}
E. Freitag, {\it Hilbert Modular Forms}, Springer-Verlag, 1990.

\bibitem{Cvetic96} M. Cveti\v{c} and D. Youm,
{\sl Rotating intersecting M-branes},
Nucl. Phys. B {\bf 499} (1997) 253; [arXiv:hep-th/9612229].

\bibitem{Waldram2}
J. P. Gauntlett, N, Kim, and D. Waldram,
{\it Supersymmetric $AdS_3, AdS_2$ and Bubble Solutions},
JHEP {\bf 0704} (2007) 005; [arXiv:hep-th/0612253].

\bibitem{Duff99}
M. J. Duff and J. T. Liu,
{\sl Anti-de Sitter black holes in gauged N=8 supergravity}, Nucl. Phys. B {\bf 554} (1999) 237; [arXiv:hep-th/9901149].

\bibitem{Bergshoeff}
E. Bergshoeff, E. Sezgin and P. Townsend,
{\sl Supermembranes and eleven-dimensional supergravity},
Phys. Lett. B {\bf 189} (1987) 75.

\bibitem{Duff91} M.J. Duff and K.S. Stelle, {\sl Multi-membrane solutions of $D=11$ supergravity}, Phys. Lett. B {\bf 253} (1991) 113.

\bibitem{Waldram1}
J. P. Gauntlett, O. A. P. Mac Conamhna, T. Mateos, and D. Waldram, {\it New supersymmetric $AdS_3$ solutions},
Phys. Rev. D {\bf 74} (2006) 106007; [arXiv:hep-th/0608055].

\bibitem{Matsushima}
Y. Matsushima and G. Shimura, {\it On the cohomology groups attached to certain vector valued differential forms on the product of the upper half planes}, Ann. of Math. {\bf 78}
(1963) 417.

\bibitem{Deligne}
P. Deligne, {\it Th\'eorie de Hodge. I, II}, Publ. Math., I.H.E.S.
{\bf 40} (1971) 5.

\bibitem{Evslin}
J. Evslin and H. Sati, {\it Can D-branes Wrap Nonrepresentable Cycles?}, JHEP {\bf 0610} (2006) 050; [arXiv:hep-th/0607045].

\bibitem{Maldacena}
J. M.~Maldacena, G. W.~Moore and N.~Seiberg,
{\it D-Brane Instantons and K-Theory Charges},
JHEP {\bf 0111} (2001) 062; [arXiv:hep-th/0108100].

\bibitem{Bouwknegt1}
P. Bouwknegt, J. Evslin and V. Mathai, {\it T-Duality: Topology Change from $H-$flux}, Commun. Math. Phys. {\bf 249} (2004) 383-415; arXiv:hep-th/0306062.

\bibitem{Bytsenko}
A. A. Bytsenko, {\it Homology and K-Theory Methods for Classes of Branes Wrapping Nontrivial Cycles}, J. Phys. A: Math. and Gen. {\bf 41} (2008) 045402; [arXiv:hep-th/0710.0305].


\bibitem{Bonora}
L. Bonora and A. A. Bytsenko, {\it Fluxes, brane charges and Chern morphisms of hyperbolic geometry}, Class. Quantum Grav. {\bf 23} (2006) 3895; [arXiv:hep-th/0602162].

\bibitem{Atiyah}
M. F. Atiyah and F. Hirzebruch, {\it Vector bundles and homogeneous spaces}, Proc. Symp. Pure Math. {\bf 3} (1961) 53.

\bibitem{Rosenberg}
J. Rosenberg, {\it Continuous trace algebras from the bundle theoretic point of view}, Jour. Aus. Math. Soc. {\bf 47} (1989) 368.

\bibitem{Bergman}
O. Bergman, E. Gimon and S. Sugimoto,
{\it Orientifolds, RR torsion, and K-theory},
JHEP {\bf 0105} (2001) 047; [arXiv:hep-th/0103183].

\bibitem{Dubrovin}
B. A. Dubrovin, A. T. Fomenko and S. P. Novikov, {\it Modern Geometry - Methods and Applications}, Part III. Introduction to Homology Theory, (Graduate Texts in Mathematics, vol. 124), Springer Verlag, 1984.

\end{thebibliography}
\end{document}